# Taxonomy of the extra-solar planet

Eva Plávalová

Dpt. of astronomy, Earth's physics, and meteorology, Comenius University, Bratislava, Slovakia
e-mail: plavala@slovanet.sk

**Abstract**

When a star is described as a spectral class G2V, we know that the star is similar to our Sun. We know its approximate mass, temperature, age, and size. When working with an extra-solar planet database, it is very useful to have a taxonomy scale (classification) such as, for example, the Harvard classification for stars. The taxonomy has to be easily interpreted and present the most relevant information about extra-solar planets. I propose the following the extra-solar planet taxonomy scale with four parameters. The first parameter concerns the mass of an extra-solar planet in the form of the units of the mass of other known planets, where M represents the mass of Mercury, E that of Earth, N Neptune, and J Jupiter. The second parameter is the planet's distance from its parent star (semi-major axis) described in logarithm with base 10. The third parameter is the mean Dyson temperature of the extra-solar planet, for which I established four main temperature classes; F represents the Freezing class, W the Water Class, G the Gaseous Class, and R the Roasters Class. I devised one additional class, however: P, the Pulsar Class, which concerns extra-solar planets orbiting pulsar stars. The fourth parameter is eccentricity. If the attributes of the surface of the extra-solar planet are known, we are able to establish this additional parameter where t represents a terrestrial planet, g a gaseous planet, and i an ice planet. According to this taxonomy scale, for example, Earth is 1E0W0t, Neptune is 1N1.5F0i, and extra-solar planet 55 Cnc e is 9E-1.8R1. Key words: Catalogues – Extra-solar planet – Habitable zone – Planets

## 1 Introduction

Astronomical papers, for the most part, specify the spectral type of stars in the first reference according to the Harvard spectral classification. When this information is known, the effective temperature of the star, its color, mass, and radius can be roughly estimated. Unfortunately, a taxonomy scale for extra-solar planets (EPs) similar to the Harvard classification for stars does not exist. We know of more than 700 EPs (as of December 2011). Moreover, Borucki et al. (2011) nominated more than 1,200 candidates for EPs from analysis of only the first four month's data. It would be helpful to investigators if a simple and comprehensive taxonomy were available when working with a large number of EPs. Some labeling for Eps has been implemented; close-in planets, for example, have been identified and so labeled, as have giant planets, Jovian planets, hot-Jupiters, hot planets, "Neptune-mass planets, Saturn-mass planets, rocky planets, and super-Earths. This labeling has contributed to the development of our knowledge of EPs and enabled investigators to identify different types of EPs simultaneously. However, a label such as "Neptune-mass planet," could refer to a close-in planet or to an ice planet as well. Other, more precise identification labels have entered into the lexicon, such as a *close-in terrestrial planet* (Haghighipour and Rastegar 2011), but by and large, a general taxonomy for EPs is yet to be established.

There are three different taxonomies in astronomy. The first was proposed by Sudarsky, Burrows, and Hubeny (2003), the second was created by Marchi (2007), and a third taxonomy was projected by Lundock et al. (2009). All these taxonomies have a benchmark in the spectrum of EPs. These are very precise taxonomies, but the initial quick data about an EP is very complicated and does not indicate the main features of EPs. Our proposal is that taxonomy data of a particular EP should be easily comparable to that of others.

## 2 Taxonomy

To establish the general features of the EP, it is requisite that, at the very least, the planet's mass and semi-major axis is known. It is helpful if the eccentricity, temperature characteristics, period, radius, and density are known as well. The seven parameters mentioned here, however, are too numerous and involved to be included in a comprehensible and comparable taxonomy. Temperature is the only parameter used in the Harvard classification. A star's mass, however, which is closely related to its temperature, is the most important condition for its evolution. For this reason only one parameter in a star's classification is necessary. Today's research indicates that, for Eps, more than one parameter for their evolution is needed.

I endeavored to choose the most important features of EPs as parameters for the proposed taxonomy. I selected the following five parameters: mass, semi-major axis, mean Dyson temperature, eccentricity, and surface attributes.

*2.1 Mass of the extra-solar planet*

It is thought that the most important parameter of an EP is its mass. The first parameter of taxonomy is information that concerns the mass of an EP. The mass of Jupiter is currently used as a benchmark mass unit for EPs. It is assumed that many more EPs will be discovered with masses less than that of Earth and possibly less than that of Mercury. I have established the units of the mass of some known planets in the Solar system. For EPs with a mass less than 0.003 $M_{Jup}$, we established the mass unit of Mercury ($3.302 \times 10^{23}$ kg). We are aware of EPs with a mass in this category. For EPs with a mass between 0.003 $M_{Jup}$ and 0.05 $M_{Jup}$, we established a mass unit scale of the Earth ($5.9736 \times 10^{24}$ kg). There are at least ten known EPs with a mass in this category. The group of EPs known as super-Earths is members of this mass unit. For EPs with a mass between 0.05 $M_{Jup}$ and 0.99 $M_{Jup}$, we established a mass unit scale of Neptune ($1.0243 \times 10^{26}$ kg). Of these EPs, there are quite a large number; more than a hundred. For the EPs with a mass more than 1 $M_{Jup}$ we used the mass unit scale of Jupiter ($1.8986 \times 10^{27}$ kg). In this category, there are currently the largest numbers of EPs.

The form of this parameter in taxonomy is the integer number of the mass unit and the first letter of the planet it corresponds to, where M represents Mercury, E Earth, N Neptune, and J Jupiter. For example, Earth is 1E, Neptune is 1N, Uranus 15E, and 55 Cnc e 9E.

*2.2 Semimajor axis*

The position of an EP in its stellar system is the next very important parameter that influences many other features of this celestial body. For this reason, the second parameter of taxonomy is the distance between the EP and its parent star in astronomical units (AU).

Initially, I had hoped to define this parameter in two different ways: one for a semi-major axis less than 1 AU different and another for a semi-major axis greater than 1 AU. For a semi-major axis less than 1 AU, I wanted to use a decimal number with one decimal position, and for a semi-major axis less than 0.1 AU, I wanted to use a decimal number with two decimal positions. For a semi-major axis greater than 1 AU, I wanted to use an integer number. However, it became clear that this method would result in a complicated and unclear outcome.

In the end, I chose to define the second parameter in logarithm form[1]. I used a logarithm with base 10 from a semi-major axis and rounded the calculated value to the nearest decimal point. For EPs with a semi-major axis smaller than 1 AU, this parameter is negative, and with the decreasing value of a semi-major axis, the value of this parameter rapidly decreases to -2. Smaller values than -2 (semi-major axis is 0.01), at present, are unexpected. For the value of a semi-major axis equal to 1 AU, the value of this parameter is 0. For a semi-major axis with a value greater than 1, the value of this parameter is positive. For example, the value of a semi-major axis equal to 10 AU has a parameter value of 1, and for a semi-major axis with a value of 100 AU, this parameter is 2.

For example, for Earth this parameter has the form of 0, for Neptune the form is 1.5, Uranus 1.3, 55 Cnc c -0.6 (the semi-major axis is 0.2403 AU), and 55 Cnc e is -1.8 (the semi-major axis is 0.0156 AU).

*2.3 Mean Dyson temperature*

The value of the surface temperature of an EP depends on many parameters, for example, albedo, speed of the rotation of the EP, or the structure of its atmosphere. A precise temperature value of an EP cannot be determined from observable data. It was necessary to establish a new universal parameter for the temperature in taxonomy, which could be determined for most known EPs.

By using the Stefan-Boltzmann law, the flux on the surface of a parent star can be expressed as

$$L = 4\pi R_*^2 T_*^4 \zeta \,, \qquad (1)$$

where $R_*$ is the radius of the star, $T_*$ the effective temperature of the star, and $\zeta$ the Stefan-Boltzmann constant. An effective radiating temperature for a planet, which is rotating slowly, can be calculated by the following equation (e.g., Karttunen et al. 2003):

$$T = T_* \left(\frac{1-A}{2}\right)^{1/4} \left(\frac{R_*}{R_{EP}}\right)^{1/2}. \qquad (2)$$

---

[1] Y. Dutil (Chaire de recherche industrielle en technologies de l'énergie et en e_cacité énergétique (T3E) École de Technologie Supérieure) private communication

Here, $R_{EP}$ is the distance from the parent star to the EP, and $A$ is Bond albedo of the planet. An effective radiating temperature for a planet that is rotating quickly is

$$T = T_* \left(\frac{1-A}{4}\right)^{1/4} \left(\frac{R_*}{R_{EP}}\right)^{1/2}. \tag{3}$$

For the EP, it can generally be said that the planet orbits quickly, and the temperature can be calculated by equation (3). Even so, it is possible that close-in planets with a very short period have a synchronous rotation, and equation (2) must be used as opposed to equation (3). This ambiguous fact leads to the establishment of a new parameter: t*he Dyson temperature*. It is a temperature that has an artificial sphere the size of a planetary orbit (Dyson sphere) (Dyson 1960), which can be defined according to the following equation.

$$T_{EP}^4 = \frac{T_*^4 R_*^2}{R_{EP}^2}. \tag{4}$$

For an EP with a small value of eccentricity, equation (4) can be used directly. However, in the case when an EP has a large value of eccentricity, the distance from the parent star changes, according to the Second Keplerian law, without homogeneity and in many cases rapidly. In some cases, the Dyson temperature changes very rapidly, too. For better precision of the calculation of the Dyson temperature, I divided the EP's orbit into ten equal segments, according to the time needed for a complete rotation. For this calculation, I used the Kepler equation. For the calculation of the eccentric anomaly, I used an iteration method with three steps (see e.g. Andrle, 1971) and calculated the value of the momentary distance of an EP from its parent star, using the following equation (see e.g. Karttunen et al. 2003):

$$R_{EP} = a(1 - e \cos E). \tag{5}$$

Here, $a$ is the semi-major axis, $e$ the eccentricity, and $E$ the eccentric anomaly of an EP. I used this value of distance in equation (4) and calculated the momentary Dyson temperature. I calculated the Dyson temperature for ten equal segments, according to the second Keplerian law, and from these ten values, I calculated the arithmetical mean, which I defined as the *mean Dyson temperature ($t_{EP}$)* for the EP.

For EPs for which the value of their albedo and speed of rotation are known, their effective radiating temperature can be calculated by

$$T = \left(\frac{1-A}{2}\right)^{1/4} t_{EP}, \tag{6}$$

for planets with a slow rotation, and by

$$T = \left(\frac{1-A}{4}\right)^{1/4} t_{EP} \qquad (7)$$

for planets with a quick rotation.

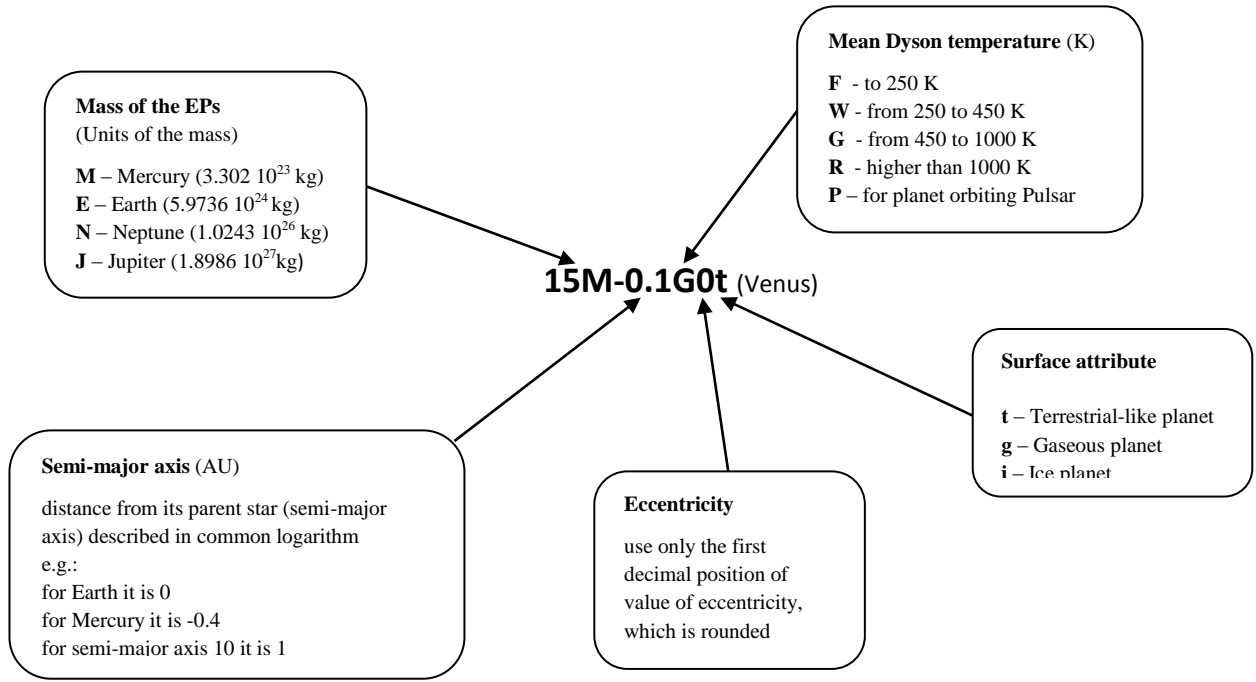

**Figure** 1: Schematic explanation - a definition of the taxonomy of the EPs for which we are able to determine the taxonomy scale.

| | | | | | |
|---|---|---|---|---|---|
| Table 1: Five EPs in system 55 Cnc (HD75732). | | | | | |
| Name | 55 Cnc b | 55 Cnc c | 55 Cnc d | 55 Cnc e | 55 Cnc f |
| Taxonomy | 15N-0.9R0 | 3N-0.6G1 | 4J0.8F0 | 9E-1.8R1 | 3N-0.1W0 |
| Mass [$M_{Jup}$] | 0.824 | 0.169 | 3.835 | 0.027 | 0.144 |
| Semi-major axis [AU] | 0.1148 | 0.2403 | 5.76 | 0.0156 | 0.781 |
| Logarithm from Semi-major axis | -0.94 | -0.62 | 0.76 | -1.81 | -0.11 |
| Mean Dyson temperature [K] | 1010.0 | 698.0 | 142.6 | 2739.4 | 387.2 |
| Eccentricity | 0.0159 | 0.053 | 0.025 | 0.057 | 0.0002 |
| Period [days] | 14.651262 | 44.3446 | 5218 | 0.73654 | 260.7 |

For example, the mean Dyson temperature of Earth is 392 K and the albedo is 0.3. With equation (7), the effective radiating temperature is 254 K.

For a clearer differentiation, I established four main temperature classes. The coldest class is the *Freezing class*, which is indicated as **F.** The members of this class have a mean Dyson temperature less than 250 K. Jupiter, Saturn, Neptune, and Uranus are members of this class. The second class is the *Water class,* and it is indicated as **W**. Members of this class have a mean Dyson temperature that ranges from 250 to 450 K. An EP from this class could have liquid water at its surface. Earth and Mars are members of this class. For the third class, the mean Dyson temperature ranges from 450 to 1000 K. This is termed the *Gaseous Class* and is indicated as **G**. Mercury and Venus are members of this class. And finally, the fourth class is the hottest class and comprised of planets with a mean Dyson temperature higher than 1000 K. This class was aptly named the *Roasters Class* and it is indicated as **R**. Many the members of this class are called close-in planets.

I defined the *Water class* as planets that reside in the Habitable zone (HZ), as defined by Kasting, Whitmire, and Reynolds (1993). These authors purported conservative estimates of HZ boundaries to be 0.95 AU for the inner edge closest to the sun and 1.37 AU for the outer edge. It should be noted that Michna et al. (2000) calculated that the outer boundary could be at distances as great as 2.4 AU, and Dole (1970) marked the inner boundaries as 0.725 AU. I used the "optimistic" boundaries 0.725 and 2.4 AU for the definition of the Water class. The value of the mean Dyson temperature for a hypothetical planet with the same parameter as Earth and orbits with a semi-major axis of 0.725 AU is 460 K. This value for a hypothetical planet with the same parameter as Earth, but with a=2.4 AU, is 253 K. Considering the fact that the boundaries of the HZ are not expressly defined, I established the range of the mean Dyson temperature for the Water class to be 250 K to 450 K.

It is known that some EPs orbit pulsar stars. For such these EPs, I predicted that their behavior would be totally different than that of EPs that orbit stars in the main sequence. I could not calculate the mean Dyson temperature of these EPs and so created an additional class, named the *Pulsar class*, which is indicated as **P**.

### 2.4 Eccentricity

As shown in the previous paragraph, the eccentricity of an EP is the next very important behavioral characteristic I included in the taxonomy as the fourth parameter. For easier reference, I used only the first decimal position for the value of eccentricity, which is rounded.

## 2.5 Surface attribute

Generally speaking, an EP's surface characteristics can not be defined; however, an improvement of observation equipment wouldallow for better and more precise data and specification of surface characteristics for many EPs in the future. For this reason, I considered a fifth additional parameter in taxonomy: the surface attribute.

I considered three different surface attributes, according to the type of surfaces that have been observed in the Solar system. The first surface attribute is *a terrestrial-like planet surface*. Mercury, Venus, Mars, and of course Earth have terrestrial-like planet surfaces, and this attribute is indicated as a lower case **t**. The second surface attribute is *a gaseous planet surface*, such as that of Jupiter and Saturn. I indicated this attribute with a lower case **g**. The third surface attribute is *an ice planet surface* similar to that of Uranus and Neptune. This attribute is indicated as a lower case **i**.

It is assumed that, as we begin to detect EPs with a mean Dyson temperature above or below that found in the Solar system, more and more EPs will emerge with different surface attributes, which will have to be defined. The possibilities are endless and may include, for example, surfaces of ocean water surface or magma. This is unheard of for us at our present level of understanding.

## 3 Interpretation examples

With all the necessary parameters defined, I considered a structure for taxonomy. Below, I present my proposed taxonomy with a sample planet from the Solar System, Venus.
The taxonomy class for Venus is **15M-0.1G0t**. That is,

**15 M** - The mass of Venus is fifteen times greater than Mercury's mass. Precisely put, the mass
of Venus is $4.8685 \times 10^{24}$ kg, roughly 15 times more than the mass of Mercury.
**-0.1** - Venus's semi-major axis is 0.723 AU. Logarithm with base 10 from this value is -0.1.
**G** - The mean Dyson temperature is between 450 K and 1000 K. The exact Dyson orbit temperature is 461 K.
**0** - Venus is orbiting on a near-circle orbit with an eccentricity value of 0.0068.
**t** - Venus has a solid surface, which we can say is a terrestrial planet surface.

The taxonomy class for the first discovered extra-solar planet orbiting a main sequence star 51 Peg b is **9N-1.3R0**. That is,

**9N** - The mass of 51Peg b is 0.468 $M_{Jup}$, which is 8.67 $M_{Nep}$ (mass of Neptune).
**-1.3** - The b semi-major axis of 51 Peg b is 0.052 AU. Logarithm with base 10 from this value is -1.3.

**Table 2**: Register for EPs orbiting one of the stars in these binary stellar systems.

| Name | 70 Vir b | eps Eridani Ab | eps Eridani Ac *(unconfirmed)* | gama Cephei b (HD 222404) | HD 142Ab | tau Boo b (HD 120136 b) |
|---|---|---|---|---|---|---|
| Taxonomy | 7J-0.3G4 | 2J0.5F7 | 2N1.6F3 | 2J0.3G0 | 1J0W4 | 4J-1.3R0 |
| Planet mass [$M_{Jup}$] | 6.6 | 1.55 | 0.1 | 1.85 | 1.03 | 3.9 |
| Semi-major axis [AU] | 0.48 | 3.39 | 40 | 2.05 | 1 | 0.046 |
| Logarithm from Semi-major axis | -0.32 | 0.53 | 1.60 | 0.31 | 0.0 | -1.34 |
| Mean Dyson temperature [K] | 737.1 | 51.6 | 174.8 | 503.2 | 385.3 | 2301.7 |
| Eccentricity | 0.43 | 0.702 | 0.3 | 0.040 | 0.37 | 0.018 |
| Period [days] | 116.67 | 2502 | 102270 | 903.3 | 339 | 3.3135 |
| Spectral class parent star | G4V | K2V | K2V | K2V | G1 IV | F7V |
| Mother star mass [$M_{Sun}$] | 0.92 | 0.83 | 0.83 | 1.4 | 1.1 | 1.3 |
| $T_{Eff}$ parent star [K] | 5432 | 5116 | 5116 | 4800 | 6180 | 6309 |
| Radius parent star [$R_{Sun}$] | 1.968 | 0.895 | 0.895 | 4.9 | 0.86 | 1.331 |

**R** - The mean Dyson temperature is higher than 1000 K; more precisely, it is 1938.6 K.
**0** - The EP 51 Peg b is orbiting on a circle or near circle orbit. The value is precisely 0.
The surface attribute parameter is mentioned only in the case when we can specify it.

The method of taxonomy as schematically explained is in Figure 1.

## 4   Practical examples

When working with large groups of EPs, it is practically impossible to compare them by using an easy and quick mechanism. For example, there are five EPs in the planetary system 55 Cnc. Without quite a wide table, it is impossible to say which EP is the smallest, which has the farthest or the closest orbit, or which planet has the most eccentric orbit, etc. Using our taxonomy, one can answer these questions practically immediately. For example, the closest planet is 55 Cnc e with a semi-major axis of about 0.02 AU (exactly written 0.0156 AU). We can immediately say that the value of the eccentricity for 55 Cnc c and 55 Cnc e is nearly the same (the exact eccentricity for 55 Cnc c is 0.053 and the eccentricity for 55 Cnc e is 0.057). The data and taxonomy identification regarding EP members of the system 55 Cnc are in Table 1.
The application of this taxonomy for EPs with a different parent star is quite significant. For example, we can compare EPs that host one of the members of the binary stellar systems. We know more than sixty systems with such a condition. The data for several EPs orbiting one of the stars in a binary stellar system are in Table 2. We only need the first line of the table to say which EP from this group is the heaviest (70 Vir b), which one is the hottest (tau Boo b), or which has the most distant orbit (eps Eridani Ac), along with many other features.
The data and taxonomy identification for a few example EPs are shown in Table 3, and the taxonomy identification for other known EPs that we can classified are in Table 4.

## 5   Conclusion

I have endeavored to build a taxonomy scale for EPs that, first, could be used as a quick and easy mechanism to determinate the main attributes for an EP and, second, allow for a quick and clear comparison of large numbers of EPs. The taxonomy scale of EPs for which we know the values for mass, semi-major axis, and eccentricity can be determined along with the values for the radius and effective temperature of their parent stars. Almost 500 known EPs have been found with this condition. For rapid determination and understanding of the taxonomy of an EP, a graphic guide is shown in Figure 1 in which the method purported here for defining the taxonomy identification of an EP is schematically interpreted

**Acknowledgments**

I am grateful to Y. Dutil and anonymous referees for their informative comments and suggestions that helped to improve the content.

Table 3: The data and taxonomy identification for a few example EPs.

| Star | spect. class | M$_*$ (M$_{Sun}$) | T$_*$ (K) | R$_*$ (R$_{Sun}$) | Planet | a$_{plan}$ | log(a$_{plan}$) | e$_{plan}$ | M$_{plan}$ (M$_{Jup}$) | Period (day) | Mean Dyson temp. (K) | Taxonomy | Ref. |
|---|---|---|---|---|---|---|---|---|---|---|---|---|---|
| CoRoT-7 | K0V | 0.93 | 5275 | 0.87 | CoRoT-7 d[u] | 0.08 | -1.097 | 0 | 0.052 | 9.021 | 1179.8 | 17E-1.1R0 | |
| eps Eridani A | K2V | 0.83 | 5116 | 0.895 | eps Eridani A c[u] | 40 | 1.602 | 0.3 | 0.1 | 102270 | 51.6 | 2N1.6F3 | |
| HAT-P-13 | G4 | 1.22 | 5638 | 1.56 | HAT-P-13 d[u] | 0.056 | -1.252 | 0 | 0.016 | 4.37 | 2018.2 | 5E-1.3R0 | |
| HD 100777 | K0 | 1 | 5582 | 1.0979[a] | HD 100777 b | 1.03 | 0.013 | 0.36 | 1.16 | 383.7 | 387.6 | 1J0W4 | 1 |
| HD 101930 | K1V | 0.74 | 5079 | 0.9259[a] | HD 101930 b | 0.302 | -0.520 | 0.11 | 0.3 | 70.46 | 602.7 | 6N-0.5G1 | 2 |
| HD 111232 | G8V | 0.78 | 5494 | 0.9187[a] | HD 111232 b | 1.97 | 0.294 | 0.2 | 6.8 | 1143 | 253.8 | 7J0.3W2 | 3 |
| HD 1461 | G0V | 1.08 | 5765 | 1.095 | HD 1461 e[u] | 1.165 | 0.066 | 0.74 | 0.072 | 454 | 373.0 | 1N0.1W7 | |
| HD 1461 | G0V | 1.08 | 5765 | 1.095 | HD 1461 d[u] | 5 | 0.699 | 0.16 | 0.3 | 5000 | 182.7 | 6N0.7F2 | |
| HD 156846 | G0V | 1.43 | 6138 | 1.9775[a] | HD 156846 b | 0.99 | -0.004 | 0.8472 | 10.45 | 359.51 | 599.0 | 10J0G8 | 4 |
| HD 16760 | G5V | 0.88 | 5620 | 0.8532[a] | HD 16760 b | 1.13 | 0.053 | 0.067 | 14.3 | 465.1 | 331.1 | 14J0.1W1 | 5 |
| HD 17092 | K0III[b] | 2.3 | 4650 | 10.1[b] | HD 17092 b | 1.29 | 0.111 | 0.166 | 4.6 | 359.9 | 880.9 | 5J0.1G2 | 6 |
| HD 17156 | G0 | 1.275 | 6079 | 1.508 | HD 17156 c[u] | 0.481 | -0.318 | 0.136 | 0.063 | 111.314 | 729.2 | 20E-0.3G1 | |
| HD 181433 | K3IV | 0.78 | 4962 | 0.75521[a] | HD 181433 d | 3 | 0.477 | 0.48 | 0.54 | 2172 | 166.4 | 10N0.5F5 | 7 |
| HD 181433 | K3IV | 0.78 | 4962 | 0.75521[a] | HD 181433 c | 1.76 | 0.246 | 0.28 | 0.64 | 962 | 219.4 | 12N0.2F3 | 7 |
| HD 181433 | K3IV | 0.78 | 4962 | 0.75521[a] | HD 181433 b | 0.08 | -1.097 | 0.396 | 0.0238 | 9.3743 | 1023.6 | 8E-1.1R4 | 7 |
| HD 181720 | G1V | 0.92 | 5781 | 1.39[b] | HD 181720 b | 1.78 | 0.250 | 0.26 | 0.37 | 956 | 345.0 | 7N0.3W3 | 8 |
| HD 190647 | G5 | 1.1 | 5628 | 1.4831[a] | HD 190647 b | 2.07 | 0.316 | 0.18 | 1.9 | 1038.1 | 322.4 | 2J0.3W2 | 1 |
| HD 208487 | G2V | 1.3 | 5929 | 1.15 | HD 208487 c[u] | 1.8 | 0.255 | 0.19 | 0.46 | 908 | 320.7 | 9N0.3W2 | |
| HD 215497 | K3V | 0.872 | 5113 | 0.8[a] | HD 215497 b | 0.047 | -1.328 | 0.016 | 0.02 | 3.93404 | 1430.7 | 6E-1.3R0 | 15 |
| HD 215497 | K3V | 0.872 | 5113 | 0.8[a] | HD 215497 c | 1.282 | 0.108 | 0.49 | 0.33 | 567.94 | 269.8 | 6N0.1W5 | 15 |
| HD 221287 | F7V | 1.25 | 6304 | 1.08237[a] | HD 221287 b | 1.25 | 0.097 | 0.08 | 3.09 | 456.1 | 397.7 | 3J0.1W1 | 1 |
| HD 2638 | G5 | 0.93 | 5192 | 0.849[a] | HD 2638 b | 0.044 | -1.357 | 0 | 0.48 | 3.4442 | 1546.8 | 9N-1.4R0 | 10 |
| HD 27894 | K2V | 0.8 | 4875 | 0.8382[a] | HD 27894 b | 0.122 | -0.914 | 0.049 | 0.62 | 17.991 | 866.5 | 11N-0.9G0 | 10 |
| HD 285968 | M2.5V | 0.49 | 3056[a] | 0.53 | HD 285968 c[u] | 0.18 | -0.745 | 0 | 0.044 | 40 | 355.7 | 14E-0.7W0 | 11 |
| HD 285968 | M2.5V | 0.49 | 3056[a] | 0.53 | HD 285968 b | 0.066 | -1.180 | 0 | 0.0265 | 8.7836 | 587.4 | 8E-1.2G0 | 11 |
| HD 330075 | G5 | 0.95 | 6295 | 0.7532[a] | HD 330075 b | 0.039 | -1.409 | 0 | 0.62 | 3.38773 | 1876.3 | 14N-1.4R0 | 12 |
| HD 38529 | G4IV | 1.48 | 5697 | 2.44 | HD 38529 d[u] | 0.74 | -0.131 | 0.23 | 0.17 | 193.9 | 699.3 | 3N-0.1G2 | |
| HD 45364 | K0V | 0.82 | 5434 | 0.8536[a] | HD 45364 c | 0.8972 | -0.047 | 0.0974 | 0.6579 | 342.85 | 359.3 | 12N0W1 | 14 |
| HD 45364 | K0V | 0.82 | 5434 | 0.8536[a] | HD 45364 b | 0.6813 | -0.167 | 0.1684 | 0.1872 | 226.93 | 411.8 | 3N-0.2W2 | 14 |
| HD 5388 | F6V | 1.21 | 6297[b] | 1.91[b] | HD 5388 b | 1.76 | 0.246 | 0.4 | 1.96 | 777 | 440.3 | 2J0.2W4 | 8 |
| HD 63454 | K4V | 0.8 | 4841 | 0.74053[a] | HD 63454 b | 0.036 | -1.444 | 0.00005 | 0.38 | 2.818049 | 1489.1 | 7N-1.4R0 | 10 |
| HD 65216 | G5V | 0.92 | 5666 | 0.9188[a] | HD 65216 b | 1.37 | 0.137 | 0.41 | 1.21 | 613.1 | 311.3 | 1J0.1W4 | 3 |
| HD 93083 | K3V | 0.7 | 4995 | 0.85679[a] | HD 93083 b | 0.477 | -0.321 | 0.14 | 0.37 | 143.58 | 453.5 | 7N-0.3G1 | 2 |
| HIP 5158 | K5V | 0.78 | 4962 | 0.45[a] | HIP 5158 c | 7.7 | 0.886 | 0.14 | 15.04 | 9018 | 81.3 | 15J0.9F1 | 15 |
| HIP 5158 | K5V | 0.78 | 4962 | 0.45[a] | HIP 5158 b | 0.89 | -0.051 | 0.54 | 1.44 | 345.63 | 235.0 | 1J-0.1F5 | 15 |
| HR 8799 | A5V | 1.56 | 7430[b] | 1.341777[a] | HR 8799 d | 27 | 1.431 | 0.1 | 10 | 41054 | 112.3 | 10J1.4F1 | 9 |
| HR 8799 | A5V | 1.56 | 7430[b] | 1.341777[a] | HR 8799 c | 42.9 | 1.632 | 0 | 10 | 82145 | 89.1 | 10J1.6F0 | 9 |
| HR 8799 | A5V | 1.56 | 7430[b] | 1.341777[a] | HR 8799 b | 68 | 1.833 | 0 | 7 | 164250 | 70.8 | 7J1.8F0 | 9 |

The data from the Extra-solar Planets Catalogue were used (Schneider et al. 2011) in this table.

a – the radius of the parent star was calculated with Stefan-Boltzmann's relation equation; b – data from reference paper; u – an unconfirmed extrasolar planet.

REFERENCES: (1) Naef et al. (2007); (2) Lovis et al. (2005); (3) Mayor et al. (2004); (4) Tamuz et al. (2008); (5) Sato et al. (2009); (6) Niedzielski et al. (2007); (7) Bouchy et al. (2009); (8) Santos et al. (2010); (9) Moya et al. (2010); (10) Moutou et al. (2005); (11) Forveille et al. (2009); (12) Pepe et al. (2004); (13) Zucker et al. (2003); (14) Correia et al. (2009); (15) Lo Curto et al. (2010).

Table 4: Taxonomy identification for known EPs (except EPs which are in the table 3).

| Planet | Taxonomy | Planet | Taxonomy | Planet | Taxonomy | Planet | Taxonomy |
|---|---|---|---|---|---|---|---|
| 11 Com b | 19J0.1R2 | GJ 581 b | 16E-1.4G0 | HD 109246 b | 14N-0.5G1 | HD 160691 e | 2J0.7F1 |
| 11 UMi b | 11J0.2R1 | GJ 581 e | 2E-1.6G3 | HD 110014 b | 11J0.3G5 | HD 16141 b | 4N-0.5G4 |
| 14 And b | 5J-0.1R8 | GJ 581 c | 5E-1.1G1 | HD 113538 b | 0N-0.1W6 | HD 16175 b | 4J0.3W6 |
| 14 Her b | 0J0.4F4 | GJ 581 d | 6E-0.7W3 | HD 113538 c | 1N0.4F3 | HD 162020 b | 14J-1.1R3 |
| 16 Cyg B b | 2J0.2W7 | Gl 179 | 15N0.4F2 | HD 114386 b | 1J0.2F2 | HD 163607 b | 14N-0.4G7 |
| 18 Del A b | 10J0.4G1 | Gl 86 b | 4J-1G0 | HD 114729 A b | 16N0.3W3 | HD 163607 c | 2J0.4W1 |
| 25 Sec c | 16N0.3G3 | Gliese 876 c | 15N-0.9W3 | HD 114762 A b | 11J-0.5G3 | HD 165409 b | 9N-0.1W3 |
| 24 Sec b | 2J0.1G1 | Gliese 876 b | 2J-0.7W0 | HD 114783 b | 1J0.1W1 | HD 164922 b | 7N0.3F1 |
| 30 Ari Bb | 10J0G3 | Gliese 876 d | 7E-1.7G2 | HD 11506 c | 15N-0.2G4 | HD 167042 b | 2J0.1G0 |
| 4 Uma b | 7J-0.1R4 | Gliese 876 e | 9N-0.5F1 | HD 11506 b | 3J0.4W2 | HD 168443 c | 17J0.5W2 |
| 42 Dra b | 3J0.1R4 | HAT-P-1 b | 10N-1.3R1 | HD 116029 b | 2J0.2G2 | HD 168443 b | 8J-0.5G5 |
| 47 Uma c | 10N0.6F1 | HAT-P-11 b | 26E-1.3R2 | HD 117207 b | 2J0.6F2 | HD 168746 b | 4N-1.2R1 |
| 47 Uma d | 2J1.1F2 | HAT-P-12 b | 4N-1.4R0 | HD 117618 b | 3N-0.8R4 | HD 1690 b | 6J0.1R6 |
| 47 Uma b | 3J0.3W0 | HAT-P-13 c | 15J0.1W7 | HD 11964 b | 12N0.5W0 | HD 169830 b | 3J-0.1G3 |
| 51 Peg b | 9N-1.3R0 | HAT-P-13 b | 16N-1.4R0 | HD 11964 c | 25E-0.6R3 | HD 169830 c | 4J0.6W3 |
| 55 Cnc A b | 15N-0.9R0 | HAT-P-14 b | 2J-1.2R1 | HD 11977 b | 7J0.3G4 | HD 170469 b | 12N0.4W1 |
| 55 Cnc A f | 3N-0.1W0 | HAT-P-15 b | 2J-1R2 | HD 121504 b | 1J-0.5G0 | HD 171028 b | 2J0.1G6 |
| 55 Cnc A c | 3N-0.6G1 | HAT-P-16 b | 4J-1.4R0 | HD 122430 b | 4J0R7 | HD 17156 b | 3J-0.8R7 |
| 55 Cnc A d | 4J0.8F0 | HAT-P-17 b | 10N-1.1R3 | HD 125612A c | 18E-1.3R3 | HD 173416 b | 3J0.1R2 |
| 55 Cnc A e | 9E-1.8R1 | HAT-P-17 c | 1J0.4F1 | HD 125612A b | 3J0.1W5 | HD 175541 b | 11N0G3 |
| 6 Lyn b | 2J0.3G1 | HAT-P-18 b | 4N-1.3R1 | HD 125612A d | 7J0.6F3 | HD 177830 b | 1J0.1G0 |
| 61 Vir c | 18E-0.7G1 | HAT-P-19 b | 5N-1.3R1 | HD 12661 b | 2J-0.1W4 | HD 177830 c | 3N-0.3G3 |
| 61 Vir d | 23E-0.3G4 | HAT-P-2 b | 9J-1.2R5 | HD 12661 c | 2J0.4W0 | HD 178911 B b | 6J-0.5G1 |
| 61 Vir b | 5E-1.3R1 | HAT-P-20 b | 7J-1.4R0 | HD 126614 b | 7N0.4W4 | HD 179079 b | 25E-1R1 |
| 70 Vir b | 7J-0.3G4 | HAT-P-21 b | 4J-1.3R2 | HD 128311 b | 2J0W3 | HD 179949 b | 18N-1.3R0 |
| alf Ari b | 2J0.1R3 | HAT-P-22 b | 2J-1.4R0 | HD 128311 c | 3J0.2F2 | HD 180314 b | 22J0.1G3 |
| BD +48 738 b | 17N0G2 | HAT-P-23 b | 2J-1.6R1 | HD 130322 b | 1J-1.1R0 | HD 180902 b | 2J0.1G1 |
| BD +48 738 b | 17N0G2 | HAT-P-24 b | 13N-1.3R1 | HD 131496 b | 2J0.3G2 | HD 181342 b | 3J0.3G2 |
| BD+20 1790 b | 7J-1.2G1 | HAT-P-25 b | 11N-1.3R0 | HD 131664 b | 18J0.5F6 | HD 183263 b | 4J0.2W4 |
| BD-10 3166 b | 9N-1.3R1 | HAT-P-26 b | 19E-1.3R1 | HD 134987 c | 15N0.8F1 | HD 183263 c | 4J0.6F3 |
| BD-17 63 b | 5J0.1F5 | HAT-P-27 b | 12N-1.4R1 | HD 134987 b | 2J-0.1G2 | HD 185269 b | 17N-1.1R3 |
| CoRoT-1 b | 1J-1.6R0 | HAT-P-28 b | 12N-1.4R1 | HD 136418 b | 2J0.1G3 | HD 187123 b | 10N-1.4R0 |
| CoRoT-10 b | 3J-1G5 | HAT-P-29 b | 14N-1.2R1 | HD 137388 b | 4N-0.1W4 | HD 187123 c | 2J0.7F3 |
| CoRoT-11 b | 2J-1.4R0 | HAT-P-3 b | 11N-1.4R0 | HD 13931 b | 2J0.7F0 | HD 18742 b | 3J0.3G2 |
| CoRoT-12 b | 1J-1.4R1 | HAT-P-30 b | 14N-1.4R0 | HD 139357 b | 10J0.4G1 | HD 188015 b | 1J0.1W2 |
| CoRoT-13 b | 1J-1.3R0 | HAT-P-31 b | 2J-1.3R2 | HD 141937 b | 10J0.2W4 | HD 189733 b | 1J-1.5R0 |
| CoRoT-14 b | 8J-1.6R0 | HAT-P-32 b | 17N-1.5R2 | HD 142 A b | 1J0W4 | HD 190228 b | 5J0.4W4 |
| CoRoT-16 b | 10N-1.2R3 | HAT-P-33 b | 14N-1.3R1 | HD 142022 A b | 5J0.5F5 | HD 190360 c | 18E-0.9R0 |
| CoRoT-17 b | 2J-1.3R0 | HAT-P-4 b | 13N-1.4R0 | HD 142245 b | 2J0.4G3 | HD 190360 b | 2J0.6F4 |
| CoRoT-18 b | 3J-1.5R1 | HAT-P-5 b | 1J-1.4R0 | HD 142415 b | 2J0W5 | HD 190984 b | 3J0.7F6 |
| CoRoT-19 b | 1J-1.3R0 | HAT-P-6 b | 1J-1.3R0 | HD 145377 b | 6J-0.3G3 | HD 192263 b | 13N-0.8G0 |
| CoRoT-2 b | 3J-1.6R0 | HAT-P-7 b | 2J-1.4R0 | HD 145457 b | 3J-0.1R1 | HD 192310 b | 17E-0.5G1 |
| CoRoT-20 b | 4J-1R6 | HAT-P-8 b | 1J-1.3R0 | HD 1461 c | 6E-1R0 | HD 192310 c | 24E0.1W3 |
| CoRoT-21 b | 3J-1.4R0 | HAT-P-9 b | 12N-1.3R0 | HD 1461 b | 8E-1.2R1 | HD 192699 b | 3J0.1G1 |
| CoRoT-23 b | 3J-1.3R2 | HD 102117 b | 3N-0.8R1 | HD 147513 b | 1J0.1W3 | HD 195019 b | 4J-0.9R0 |
| CoRoT-3 b | 22J-1.2R0 | HD 102195 b | 8N-1.3R0 | HD 148156 b | 16N0.4W5 | HD 196050 b | 3J0.4W2 |
| CoRoT-4 b | 13N-1R0 | HD 102272 c | 3J0.2G7 | HD 148427 b | 18N0G2 | HD 196885 A b | 3J0.4W5 |
| CoRoT-5 b | 9N-1.3R1 | HD 102272 b | 6J-0.2R1 | HD 149026 b | 7N-1.4R0 | HD 19994 b | 2J0.2G3 |
| CoRoT-6 b | 3J-1.1R1 | HD 102329 b | 6J0.3G2 | HD 1502 b | 3J0.1G1 | HD 200964 c | 1J0.3G2 |
| CoRoT-7 b | 5E-1.8R0 | HD 102956 b | 1J-1.1R0 | HD 152581 b | 2J0.2G2 | HD 200964 b | 2J0.2G0 |
| CoRoT-7 c | 8E-1.3R0 | HD 104985 b | 6J-0.1R0 | HD 153950 b | 3J0.1W3 | HD 202206 b | 17J-0.1W4 |
| CoRoT-8 b | 4N-1.2R0 | HD 106252 b | 8J0.4F5 | HD 154672 b | 5J-0.2G6 | HD 202206 c | 2J0.4F3 |
| CoRoT-9 b | 16N-0.4G1 | HD 106270 b | 11J0.6W4 | HD 154857 b | 2J0.1G5 | HD 20367 b | 1J0.1W2 |
| eps Eridani A b | 2J0.5F7 | HD 10647 b | 17N0.3W1 | HD 156411 b | 14N0.3W2 | HD 2039 b | 5J0.3W7 |
| Fomalhaut b | 3J2.1F1 | HD 10697 b | 6J0.3W1 | HD 156668 b | 4E-1.3R0 | HD 204941 b | 5N0.4F4 |
| gam 1 Leo A b | 9J0.1R1 | HD 108147 b | 5N-1R5 | HD 158038 b | 2J0.2G3 | HD 205739 b | 1J0G3 |
| gam Ceph A b | 2J0.3G0 | HD 108863 b | 3J0.1G1 | HD 160691 d | 10N0W1 | HD 206610 b | 2J0.2G2 |
| GJ 1214 b | 6E-1.9G3 | HD 108874 c | 1J0.4F3 | HD 160691 c | 11E-1R2 | HD 208487 b | 8N-0.3G2 |
| GJ 436 b | 23E-1.5R2 | HD 108874 b | 1J0W1 | HD 160691 b | 2J0.2W1 | HD 20868 b | 2J0W8 |

Table 4: continued.

| Planet | Taxonomy | Planet | Taxonomy | Planet | Taxonomy | Planet | Taxonomy |
|---|---|---|---|---|---|---|---|
| HD 209458 b | 13N-1.3R0 | HD 47536 b | 5J0.2R2 | HD 98219 b | 2J0.1G2 | Saturn | 6N1F1g |
| HD 210277 b | 1J0W5 | HD 49674 b | 2N-1.2R2 | HD 99492 b | 2N-0.9G3 | tau Boo b | 4J-1.3R0 |
| HD 210702 b | 2J0.1G2 | HD 50499 b | 2J0.6F2 | HD 99492 c | 7N0.7F1 | TrES-2 b | 1J-1.4R0 |
| HD 212771 b | 2J0.1G1 | HD 50554 b | 5J0.4W5 | HD 99706 b | 1J0.3G4 | TrES-3 b | 2J-1.6R0 |
| HD 213240 b | 5J0.3W5 | HD 52265 b | 1J-0.3G4 | HIP 13044 b | 1J-0.9R3 | TrES-4 b | 17N-1.3R0 |
| HD 216435 b | 1J0.4W1 | HD 5319 b | 2J0.2G1 | HIP 14810 d | 11N0.3W2 | ups And d | 10J0.4W3 |
| HD 216437 b | 2J0.4W3 | HD 5891 b | 8J-0.1R1 | HIP 14810 c | 1J-0.3G2 | ups And b | 13N-1.2R0 |
| HD 216770 b | 12N-0.3G4 | HD 62509 b | 3J0.2G0 | HIP 14810 b | 4J-1.2R1 | ups And c | 15J-0.1G2 |
| HD 217107 b | 1J-1.1R1 | HD 6434 b | 7N-0.9G2 | HIP 57050 b | 0N-0.8W3 | ups And e | 1J0.7F0 |
| HD 217107 c | 2J0.7F5 | HD 6718 b | 2J0.6F1 | HIP 57274 d | 10N0W3 | V391 Peg b | 3J0.2G0 |
| HD 217786 b | 13J0.4W4 | HD 68988 b | 2J-1.1R1 | HIP 57274 b | 11E-1.2G2 | WASP-1 b | 16N-1.4R0 |
| HD 218566 b | 4N-0.2W3 | HD 69830 b | 10E-1.1R1 | HIP 57274 c | 8N-0.7G1 | WASP-10 b | 3J-1.4R1 |
| HD 219828 b | 21E-1.3R0 | HD 69830 c | 12E-0.7G1 | HIP 75458 b | 9J0.1G7 | WASP-11 b | 9N-1.4R0 |
| HD 222582 b | 8J0.1W7 | HD 69830 d | 18E-0.2W1 | HR 810 b | 2J0G2 | WASP-12 b | 1J-1.6R0 |
| HD 224693 b | 13N-0.6R1 | HD 70642 b | 2J0.5F1 | kappa CrB b | 2J0.4W2 | WASP-14 b | 8J-1.4R1 |
| HD 23079 b | 3J0.2W0 | HD 7199 b | 5N0.1W2 | Kepler-10 b | 0E-1.8R0 | WASP-15 b | 10N-0.3G0 |
| HD 231701 b | 1J-0.3G1 | HD 72659 b | 3J0.7F2 | Kepler-10 c | 0N-0.6G0 | WASP-16 b | 16N-1.4R0 |
| HD 23596 b | 8J0.5W3 | HD 73256 b | 2J-1.4R0 | Kepler-11 c | 14E-1R0 | WASP-17 b | 9N-1.3R0 |
| HD 240237 b | 5J0.3R4 | HD 73267 b | 3J0.3F3 | Kepler-11 g | 18N-0.3G0 | WASP-18 b | 10J-1.7R0 |
| HD 25171 b | 1N0.5W1 | HD 73526 b | 3J-0.2G2 | Kepler-11 b | 4E-1R0 | WASP-19 b | 1J-1.8R0 |
| HD 27442 b | 1J0.1G1 | HD 73526 c | 3J0G1 | Kepler-11 f | 4M-0.6G0 | WASP-2 b | 16N-1.5R0 |
| HD 28185 b | 6J0W1 | HD 73534 b | 1J0.5W0 | Kepler-11 d | 6E-0.8R0 | WASP-21 b | 6N-1.3R0 |
| HD 28254 b | 1J0.3W8 | HD 74156 b | 2J-0.5G6 | Kepler-11 e | 8E-0.7G0 | WASP-22 b | 1N-1.3R0 |
| HD 28678 b | 2J0.1G2 | HD 74156 d | 7N0G3 | Kepler-12 b | 8N-1.3R0 | WASP-23 b | 16N-1.4R1 |
| HD 290327 b | 3J0.5F1 | HD 74156 c | 8J0.5W4 | Kepler-16(AB) b | 6N-0.2W0 | WASP-24 b | 1J-1.4R0 |
| HD30177 b | 8J0.4F2 | HD 7449 b | 1J0.4W8 | Kepler-17 b | 2J-1.6R0 | WASP-25 b | 1N-1.3R0 |
| HD 30562 b | 1J0.4W6 | HD 7449 c | 2J0.7F5 | Kepler-4 b | 24E-1.3R0 | WASP-26 b | 1J-1.4R0 |
| HD 30856 b | 2J0.3G2 | HD 75289 b | 8N-1.3R0 | Kepler-5 b | 2J-1.3R0 | WASP-28 b | 1N-1.3R0 |
| HD 31253 b | 1N0.1G3 | HD 75898 b | 3J0.1G1 | Kepler-6 b | 12N-1.3R0 | WASP-29 b | 5N-1.3R0 |
| HD 32518 b | 3J-0.2R0 | HD 76700 b | 4N-1.3R0 | Kepler-7 b | 8N-1.2R1 | WASP-3 b | 2J-1.5R0 |
| HD 33142 b | 1J0G2 | HD 7924 b | 9E-1.2R2 | Kepler-8 b | 11N-1.3R0 | WASP-31 b | 0N-1.3R0 |
| HD 33283 b | 6N-0.8R5 | HD 80606 b | 4J-0.3G9 | KOI-196 b | 9N-1.5R0 | WASP-32 b | 4J-1.4R0 |
| HD 33564 b | 9J0W3 | HD 81040 b | 7J0.3F5 | KOI-423 b | 18J-0.8R1 | WASP-33b | 5J-1.6R0 |
| HD 34445 b | 15N0.3W3 | HD 81688 b | 3J-0.1R0 | KOI-428 b | 2J-1.1R0 | WASP-34 b | 11N-1.3R0 |
| HD 3651 b | 4N-0.5G6 | HD 82886 b | 1J0.2G3 | ksi Aql b | 3J-0.2R0 | WASP-36 b | 2J-1.6R0 |
| HD 37124 c | 12N0.2W1 | HD 82943 c | 2J-0.1G4 | Lupus-TR-3 b | 15N-1.3R0 | WASP-37 b | 2J-1.4R0 |
| HD 37124 b | 13N-0.3G0 | HD 82943 b | 2J0.1W2 | OGLE-TR-10 b | 13N-1.4R0 | WASP-39 | 5N-1.3R0 |
| HD 37124 d | 13N0.4F2 | HD 83443 b | 8N-1.4R0 | OGLE-TR-111 b | 10N-1.3R0 | WASP-4 b | 1J-1.6R0 |
| HD 38529 b | 14N-0.9R2 | HD 8535 b | 13N0.4W2 | OGLE-TR-113 b | 1J-1.6R0 | WASP-41 b | 17N-1.4R0 |
| HD 38529 c | 18J0.6W4 | HD 8574 b | 2J-0.1G3 | OGLE-TR-132 b | 1J-1.5R0 | WASP-43 b | 2J-1.8R0 |
| HD 38801 b | 11J0.2W0 | HD 86081b | 2J-1.4R0 | OGLE-TR-182 b | 1J-1.3R0 | WASP-44 b | 1J-1.5R0 |
| HD 39091 b | 10J0.5W6 | HD 86264 b | 7J0.5W7 | OGLE-TR-211 b | 14N-1.3R0 | WASP-45 b | 1J-1.4R0 |
| HD 40979 b | 3J-0.1G3 | HD 8673 b | 14J0.5W7 | PRS 1257+12 b | 0M-0.7P0 | WASP-46 b | 2J-1.6R0 |
| HD 41004 A b | 3J0.2W4 | HD 87883 b | 12J0.6F5 | PRS 1257+12 d | 4E-0.3P0 | WASP-48 b | 18N-1.5R0 |
| HD 4113 b | 2J0.1W9 | HD 88133 b | 4N-1.3R1 | PRS 1257+12 c | 4E-0.4P0 | WASP-5 b | 2J-1.6R0 |
| HD 4203 b | 2J0.1W5 | HD 89307 b | 2J0.5F2 | PSR 1719-14 b | 1J-3.4P1 | WASP-50 b | 1J-1.5R0 |
| HD 4208 b | 15N0.2W0 | HD 89744 b | 7J-0.1G7 | Qatar-1 b | 1J-1.6R0 | WASP-51 b | 14N-1.4R0 |
| HD 4308 b | 13E-0.9R3 | HD 92788 b | 4J0W3 | rho CrB b | 1J-0.7G0 | WASP-6 b | 9N-1.4R1 |
| HD 4313 b | 2J0.1G0 | HD 9446 b | 13N-0.7G2 | Venus | 15M-0.1G0t | WASP-7 b | 18N-1.2R0 |
| HD 43197 b | 11N0W8 | HD 9446 c | 2J-0.2G1 | Uranus | 15E1.3F0i | WASP-8 b | 2J-1.1R3 |
| HD 44219 b | 11N0.1W6 | HD 95089 b | 1J0.2G2 | Earth | 1E0W0t | XO-2 b | 11N-1.4R0 |
| HD 45350 b | 2J0.3W8 | HD 96063 b | 17N0G3 | Jupiter | 1J0.7F0g | XO-3 b | 12J-1.3R3 |
| HD 46375 b | 5N-1.4R0 | HD 96127 b | 4J0.1R3 | Mercury | 1M-0.4G2t | XO-4 b | 2J-1.3R0 |
| HD 47186 b | 23E-1.3R0 | HD 96167 b | 13N0.1G7 | Neptune | 1N1.5F0i | XO-5 b | 1J-1.3R0 |
| HD 47186 c | 7N0.4W2 | HD 97658 b | 6E-1.1R1 | Mars | 2M0.2W1t | | |

The data from the Extra-solar Planets Catalogue were used (Schneider et al. 2011) in this table.